\documentclass[sigconf]{acmart}
\usepackage{adjustbox}
\usepackage[ruled,vlined]{algorithm2e}
\usepackage{comment}
\usepackage{xcolor}
\usepackage[table]{xcolor}
\usepackage{soul}
\usepackage{subcaption}

\SetCommentSty{mycommentstyle}

\AtBeginDocument{%
  }

\makeatletter
\def\@specialsection#1{%
  \let\@vspace\@vspace@orig
  \let\@vspacer\@vspacer@orig
  \ifcase\ACM@format@nr
    \par\medskip\noindent#1: %
  \or
    \par\medskip\noindent#1: %
  \or
    \par\medskip\noindent#1: %
  \or
    \par\medskip\noindent#1: %
  \or 
    \par\smallskip\noindent\textbf{#1}\par\nobreak\smallskip
  \fi
  \let\@vspace\@vspace@acm
  \let\@vspacer\@vspacer@acm
}
\makeatother
\setcopyright{acmlicensed}
\copyrightyear{2026}
\acmYear{2026}
\setcopyright{cc}
\setcctype{by}
\acmConference[SIGIR '26]{Proceedings of the 49th International ACM SIGIR Conference on Research and Development in Information Retrieval}{July 20--24, 2026}{Melbourne, VIC, Australia}
\acmBooktitle{Proceedings of the 49th International ACM SIGIR Conference on Research and Development in Information Retrieval (SIGIR '26), July 20--24, 2026, Melbourne, VIC, Australia}
\acmDOI{10.1145/3805712.3809682}
\acmISBN{979-8-4007-2599-9/2026/07}

\usepackage{multirow}
\usepackage{tikz}
\usepackage{subcaption}
\usepackage{xcolor}
\usepackage{colortbl}
\usepackage{tabularx}
\usepackage{makecell}
\usepackage{array}
\usepackage{csquotes}
\usepackage{makecell}
\usepackage{arydshln}
\usepackage{pdfpages}
\usepackage{graphicx}
\usepackage{comment}
\usepackage{xspace}
\usepackage{enumitem}
\usepackage{color}

\newcommand{\algname}{MUDY}

\begin{document}

\title{\algname: Multi-Granular Dynamic Candidate Contextualization for Unsupervised Keyphrase Extraction }
\author{Hyeongu Kang}
\email{hyeongu_kang@korea.ac.kr}
\affiliation{%
  \institution{Korea University}
  \city{Seoul}
  \country{Republic of Korea}
}

\author{Susik Yoon}
\email{susik@korea.ac.kr}
\affiliation{%
  \institution{Korea University}
  \city{Seoul}
  \country{Republic of Korea}
}

\renewcommand{\shortauthors}{Hyeongu Kang et al.}

\begin{abstract}
Keyphrase extraction aims to automatically identify concise phrases that effectively represent the content of a document. While recent methods leveraging pre-trained language models (PLMs) have significantly improved the extraction of keyphrases with strong global semantic relevance, they often fall short in capturing the local contextual importance of keyphrases tied to specific subtopics dispersed in a document.
In this paper, we propose a novel context-centric framework, \algname{}, that effectively captures multi-granular contextual salience of candidate keyphrases. \algname{} employs two complementary components: (1) a prompt-based scoring that estimates the generation likelihood of each candidate keyphrase, augmented with candidate-aware weighting to better reflect its local contextual importance, and (2) a self-attention-based scoring that utilizes multi-granular attention patterns from PLMs to assess candidate significance at both the document-wide and segment-specific levels.
Evaluations on four real-world datasets demonstrate that \algname{} outperforms state-of-the-art baselines in top-k accuracy at various cutoff thresholds. In-depth quantitative and qualitative analyses further highlight the efficacy of context-centric keyphrase extraction with multi-granular saliency. For reproducibility, the source code of \algname{} is available at \url{https://github.com/HgKang1/MUDY}.
\end{abstract}



\maketitle
\vspace*{-10pt}
\bibliographystyle{ACM-Reference-Format}

\section{Introduction}

\begin{figure}[t]
  \centering
  \includegraphics[width=0.43\textwidth]{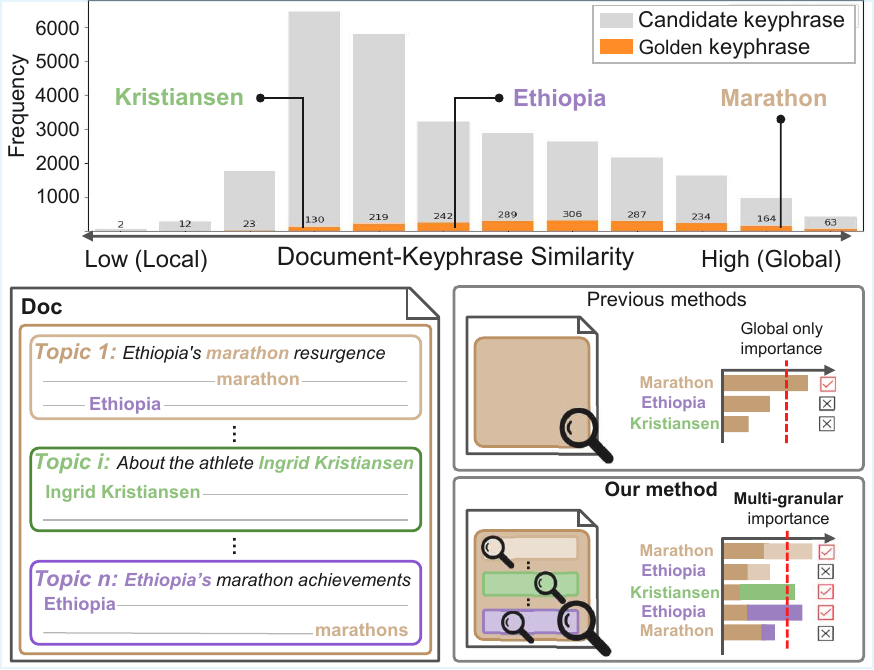}
  \vspace{-0.2cm}
  \caption{
    The document-keyphrase similarity distribution, with an example illustrating keyphrase scoring~\cite{duc}. Golden keyphrases span a broad range of similarities, revealing the difficulty of capturing multi-granularly salient keyphrases. 
  }
  \label{fig:motivation1}
    \vspace{-0.6cm}
\end{figure}

Keyphrase extraction aims to identify phrases that represent a document's core concepts and diverse subtopics to provide a granular and comprehensive summary. It is widely used for various information retrieval tasks, such as summarization~\cite{downstream_story, textsum}, retrieval-augmented generation~\cite{downstream_rag}, and topic modeling ~\cite{downstream_topicmodel, downstream_topictaxo}. \textit{Unsupervised} keyphrase extraction is particularly attractive for its broad applicability, as they do not require labeled training data and can be directly applied to new corpora across diverse domains.

Traditional unsupervised keyphrase extraction methods fall into two main categories: statistics-based and graph-based approaches. Statistics-based methods~\cite{tf-idf,yake} typically rely on simple features such as word frequency or positional information. Graph-based methods ~\cite{textrank, duc, topicrank, multirank} construct co-occurrence graphs within a document and rank keyphrases based on graph centrality. Although these methods are computationally efficient, they often fail to capture deep semantics, leading to suboptimal extraction results. 

With the advancement of pre-trained language models (PLMs), three major categories of semantic-aware keyphrase extraction have emerged. First, embedding-based methods~\cite{embedrank,jointgl,mderank} prioritize keyphrase candidates based on their semantic similarity to the document in the contextualized embedding space. Second, attention-based methods~\cite{samrank,attention-seeker} utilize the self-attention map from PLMs to identify candidates with high attention scores across all tokens. Third, prompt-based methods ~\cite{kong2023promptrank, empirical} estimate candidates based on generation probabilities derived by a PLM and a given prompt.

Existing methods, however, primarily model the semantic relationship between the \textit{entire} document and individual candidates, often overlooking the local subtopics. Since real-world documents typically include interrelated subtopics that collectively convey the document's comprehensive ideas, golden keyphrases encompass \textit{multi-granular topics} reflecting both global and local themes, as exemplified in Figure~\ref{fig:motivation1}. Consequently, identifying keyphrases that represent these multi-granularly salient topics is vital for enhancing document understanding and facilitating downstream tasks.

Despite this necessity, existing methods exhibit fundamental limitations in capturing the diverse contextual importance of individual candidates. Most existing approaches model the relevance of keyphrases to the entire document context through semantic similarity (embedding-based), aggregated attention scores (attention-based), or conditioned generation probabilities (prompt-based), treating all tokens as equally important. While JointGL~\cite{jointgl} attempts to incorporate additional context, its reliance on a semantic similarity graph and a static positional bias favoring document boundaries limits its ability to capture how a candidate's importance varies across its distinct occurrences within the document. 

Effectively capturing multi-granular topic saliency is essential for extracting keyphrases that are emphasized across varying contexts. As illustrated in Figure~\ref{fig:motivation1}, the state-of-the-art prompt-based method~\cite{kong2023promptrank} successfully identifies the global keyphrase \texttt{Marathon}, yet fails to retrieve \texttt{Ethiopia} and \texttt{Kristiansen}. These latter terms are salient only within specific subtopics dispersed throughout the document. Accounting for such multi-granular saliency is crucial for effective and practical keyphrase extraction, which is the primary objective of this work.

The core idea to capture the varying saliency of candidates is to pivot from a traditional position-centric view to a \textit{context-centric view}, inspired by Lexical Chain theory in linguistics~\cite{lexical_chain}. This theory posits that the contextual meaning of a word is shaped by both its immediate surrounding neighbors and related contexts distributed across the document. Rather than relying on static positional priors~\cite{jointgl, kong2023promptrank}, we evaluate a candidate's importance based on its specific local contexts. Recognizing that a candidate with the same surface form may hold varying degrees of saliency depending on the underlying subtopic, we conduct an instance-level evaluation of importance for candidates across their distinct occurrences.

To this end, we introduce an unsupervised framework \textbf{\algname{}}, exploiting \underline{MU}lti-granular \underline{DY}namic contextualization for keyphrase extraction, by adopting a prompt-based approach to leverage the power of PLMs. We observe that context-centric evaluation of individual candidate occurrence within PLMs is not straightforward. PLMs' attention mechanisms often suffer from attention dispersion, where the model's focus is diluted over irrelevant tokens~\cite{att_map_flat}, thereby failing to capture the precise local importance of specific keyphrase mentions. To overcome this, \algname{} applies a \textit{candidate-aware weighting} mechanism to prompt-document attention maps with a layer-wise scaling strategy. This ensures that the generation process is explicitly guided by the immediate local context of each candidate. Furthermore, to mitigate the linguistic bias of PLMs toward high-frequency or generic patterns, \algname{} employs a \textit{multi-granular attention} mechanism. By segmenting documents into overlapping windows, \algname{} computes local attention scores that are dynamically integrated with global semantic patterns. This complementary dual-stream approach, combining generative semantic richness with multi-granular contextual saliency, allows for robust, instance-level ranking of keyphrases under fine-grained specific themes as well as a broad document-level theme.

In brief, \algname{} is an unsupervised keyphrase extraction framework that leverages multi-granular topic saliency. By utilizing candidate-aware weighting and multi-granular attention, it models the instance-level local contexts of candidates, motivated by the lexical chain. Extensive experiments demonstrate that integrating global context with enhanced local saliency leads to improved extraction performance. We evaluate \algname{} against state-of-the-art models on four real-world datasets, showing strong and consistent gains, particularly on long documents and those with high topic drift. Our main contributions are summarized as follows:
\begin{itemize}[leftmargin=10pt, noitemsep]
    \item We propose \algname{}, which moves beyond static positional biases to instance-level dynamic contextual scoring to effectively capture multi-granular topic saliency.
    \item We demonstrate the effectiveness of \algname{} for long documents with high topical diversity, while remaining competitive on short and topically coherent corpora, with widely used PLMs for a prompt-based approach such as encoder-decoder models (e.g., T5) and decoder-only models (e.g., Gemma2).
    \item We conduct an in-depth evaluation of \algname{} on four real-world datasets. \algname{} achieves average performance gains of 6.52\%, 5.59\%, and 5.35\% over 12 baselines at F1@5, F1@10, and F1@15, respectively. Efficacy of key components and practical merits are demonstrated with quantitative and qualitative analyses.
\end{itemize}

\section{Related Work}

\vspace{0.1cm}
\noindent\emph{\underline{Unsupervised Keyphrase Extraction}}: 
Unsupervised keyphrase extraction methods are typically categorized into statistical and graph-based approaches. Statistical methods assess candidate importance using features like word frequency and position, as in YAKE~\cite{yake}. Graph-based methods, such as TextRank~\cite{textrank}, model candidates as graph nodes linked by co-occurrence and ranked with PageRank. Extensions include SingleRank~\cite{duc}, TopicRank~\cite{topicrank}, and MultipartiteRank~\cite{multirank}. Recent advances in PLMs have led to methods that leverage semantic representations. 

\noindent\emph{\underline{Embedding-based Methods}}:
Embedding-based methods rank candidates by their similarity to the document embedding, as in EmbedRank~\cite{embedrank} using Doc2Vec~\cite{d2v} and Sent2Vec~\cite{s2v}. MDERank~\cite{mderank} addresses the length mismatch problem between documents and candidate phrases by measuring the similarity gap between the original document and the document with the candidate masked. JointGL~\cite{jointgl} is a notable method that jointly models global and local information, but its notion of locality mainly relies on semantic similarity among candidates and a positional bias toward document boundaries. Consequently, JointGL's local scoring primarily models static word-to-word relations and global position, falling short in representing the semantic flow and structural coherence of the surrounding text. Our method instead adopts a context-centric view of locality to better capture multi-granular topical structures.

\noindent\emph{\underline{Attention-based Methods}}: 
Self-attention is central to Transformer models, enabling tokens to capture contextual dependencies via the self-attention map (SAM), an alignment of Query and Key vectors. SAMs provide interpretable insights into token interactions, and prior work~\cite{bertsam, attention} has shown that patterns vary across layers and heads. Based on these observations, SAMRank~\cite{samrank} identifies globally important tokens by aggregating attention weights each token receives from all others via column-wise summation of selected SAMs. However, it relies solely on global attention and aggregates across all occurrences of a candidate, which biases the model toward frequent terms and overlooks locally salient keyphrases. To address this, we adopt occurrence-aware scoring that distinguishes multiple mentions of the same candidate by combining global attention with local attention computed from overlapping sentence windows. The two occurrence-aware scores are then aggregated, allowing for adaptive scoring of candidate mentions.

\noindent\emph{\underline{Prompt-based Methods}}: 
Prompt learning has emerged as a paradigm in NLP ~\cite{prompt-learning,prompt_learning1,prompt_learning2}, offering an alternative to fine-tuning by leveraging implicit knowledge in PLMs. Instead of updating parameters, these approaches perform downstream tasks by conditioning on task-specific templates. PromptRank ~\cite{kong2023promptrank} was the first method to adopt prompt learning for unsupervised keyphrase extraction. It employs the encoder-decoder architecture of PLMs to estimate the generation probability of each candidate using a template such as “\textit{This book mainly talks about} [candidate].” 
Empirical~\cite{empirical} investigates zero-shot keyphrase extraction using decoder-only PLMs and shows that diverse prompting strategies, particularly role prompting, significantly influence keyphrase extraction performance.

While PromptRank demonstrates the potential of prompt-based generation for keyphrase extraction, its scoring based on cross-attention between the template and document tokens is often broadly distributed, diluting key local contextual signals. As a result, PromptRank struggles to capture the importance of candidates within specific sections or subtopics.
To address this limitation, we apply candidate-aware weighting on the template-document map, emphasizing the local context centered around each candidate. Additionally, we incorporate the local context of semantically similar words via a Gaussian mixture weighting scheme and select the top document tokens based on their attention scores with respect to the candidate, enhancing the sensitivity to relevant local cues.


\section{Preliminaries}


\vspace{0.1cm}
\noindent\emph{\underline{Prompt-based Candidate Scoring}}:
Following PromptRank ~\cite{kong2023promptrank}, we compute a generation score for each candidate by inserting it into a predefined template and measuring its length-normalized log-likelihood, a widely used metric for sequence probability ~\cite{log-likeli1}. Since longer sequences tend to have lower total log-probabilities, normalization ensures fair comparison across different candidate lengths:
\begin{equation}
\label{eq:gen_prob}
p_c = \frac{1}{\lvert c \rvert^\alpha} \sum_{i=k}^{k+\lvert c \rvert - 1} \log p(y_i \mid y_{<i}),
\end{equation}
where $k$ is the start index of the candidate $c$ in the template, $\lvert c \rvert$ is the length of the candidate $c$, and $\alpha$ is a hyperparameter controlling the length penalty. Note that since $p_c$ is a length-normalized log-likelihood, its value is typically negative. We therefore rank candidates in descending order of $p_c$, such that higher (i.e., less negative) values indicate greater importance.


\noindent\emph{\underline{PLM Architectures}}: 
We evaluate candidates using decoder logits derived from two main types of PLMs: encoder-decoder (e.g., T5~\cite{t5}) and decoder-only (e.g., Gemma2~\cite{gemma2}) architectures. In encoder-decoder models, the document is first encoded into contextual embeddings, and the decoder uses a prompt containing the candidate to estimate its likelihood based on both the encoded input and the prompt context. On the other hand, decoder-only models treat the document and prompt as a single, continuous input sequence, processing it autoregressively. These models compute the probability of each token in the candidate based on all prior tokens, relying solely on self-attention without cross-attention.


\section{Methodology}

\begin{figure*}[t]
    \centering
    \subfloat[Overall procedure of \algname{}.]{%
        \includegraphics[height=6.1cm, keepaspectratio]{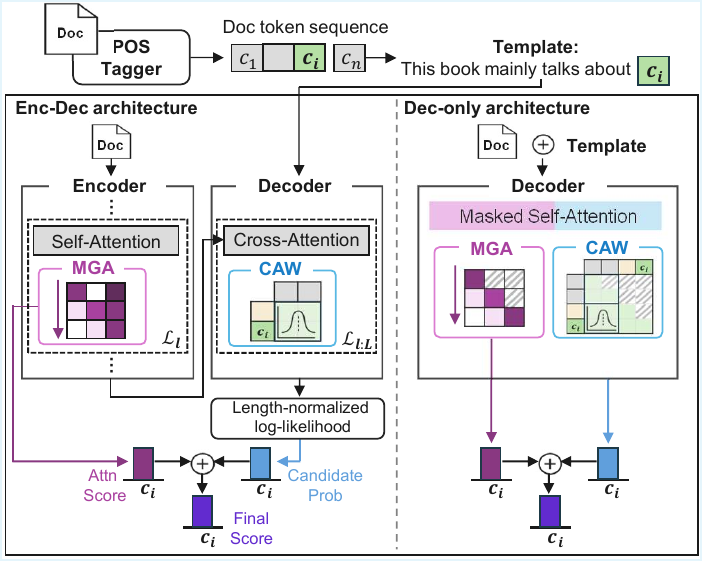}
        \label{fig:overview-a}
    }
    \hfill
    \subfloat[Candidate-aware weighting.]{%
        \includegraphics[height=6.1cm, keepaspectratio]{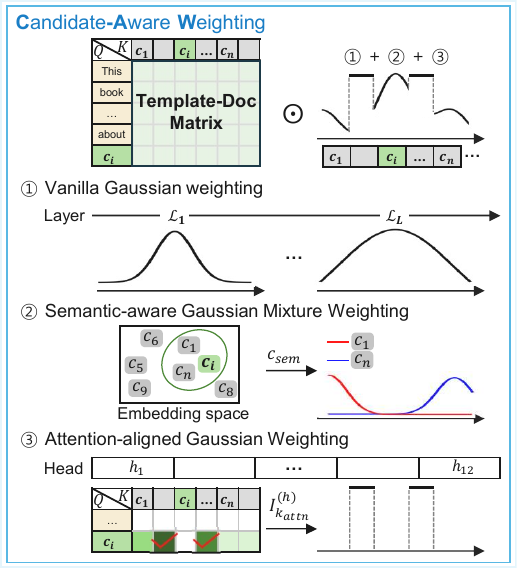}
        \label{fig:overview-b}
    }
    \hfill
    \subfloat[Multi-granular attention.]{%
        \includegraphics[height=6.1cm, keepaspectratio]{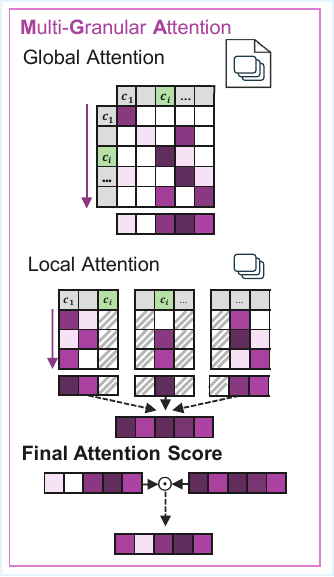}
        \label{fig:overview-c}
    }
    \vspace{-0.2cm}
    \caption{The overview of \algname{}. Two main components, candidate-aware weighting and multi-granular attention, are applied to encoder-decoder or decoder-only models. The combined scores from the two components are used to rank keyphrases.}
    \label{fig:overview}
\end{figure*}

\begin{algorithm}[t]
\small
\LinesNumbered  
\caption{Overall Procedure of \algname{}}
\label{alg:pseudo_code}
\KwIn{Document $d$, Overlapping blocks $\mathcal{B}$, model $f$}
\KwOut{Top-$N$ Keyphrase Candidates}

\tcc{Candidates generation (Sec 4.1)}
$C \gets \texttt{GenerateCandidates}(d)$

\tcc{Candidate-aware weighting (Sec 4.2)}
\For{each candidate $c \in C$}{
    \For{each layer, head $h$}{
        $S^{(h)} \gets \text{InternalCrossAttn}(f, d, c)$ 
        $W^{(h)}_{final,c} \gets \texttt{BuildCAW}(c, S^{(h)})$
        $A^{(h)} \gets \texttt{softmax}(\frac{S^{(h)} \odot W^{(h)}_{final,c}}{\sqrt{d_N}})$
    }
    $p_c \gets \text{ForwardWithModifiedAttention}(f \mid \{A^{(h)}\})$
}

\tcc{Multi-granular attention (Sec 4.3)}
$\mathbf{A} \gets \texttt{ExtractSAM}(d, f)$  
$\{G_c\}_{c\in C} \gets \texttt{ComputeGlobalScores}(C, \mathbf{A})$

\For{each block $b_k \in \mathcal{B}$}{
    $A^{\text{local}}_{b_k} \gets \texttt{ExtractSAM}(b_k, f)$
    \For{each candidate $c \in$ range of $b_k$}{
        $L_c \gets \texttt{ComputeLocalScore}(c, A^{\text{local}}_{b_k})$
    }
}

\For{each candidate $c \in C$}{
    $AS_c \gets G_c \times L_c$
}

\tcc{Final candidate scoring (Sec 4.4)}
\For{each candidate $c \in C$}{
    $FS_c \gets \frac{p_c}{1 + \log(1 + \lambda \cdot AS_c)}$
}

\KwRet{Top-$N$ candidates by $FS_c$}
\end{algorithm}

In this section, we introduce our proposed method \algname{} in detail. Figure~\ref{fig:overview} and Algorithm~\ref{alg:pseudo_code} illustrate the core architecture and overall procedure of our method, which consists of four main steps: (1) Generate a candidate set  $C = \{ c_1, c_2, \dots, c_n \}$ from the given document $d$ using part-of-speech sequences. (2) Calculate the generation probability $p_c$ for each candidate $c \in C$ using a prompt-based approach with our proposed local-context-aware cross-attention weighting. (3) Compute the attention-based importance score $AS_c$ by combining the global attention score obtained from the self-attention map (SAM) over the entire document and the local attention score, calculated from the SAM within the overlapping window. (4) Integrate the generation probabilities and SAM-based scores to obtain the final importance scores for each candidate occurrence, and rank all candidates accordingly.

\subsection{Candidates Generation}
We follow the common protocol to generate candidates. Specifically, we tokenize all words in the document, perform POS tagging, and extract noun phrases as keyphrase candidates using the regular expression pattern <NN.*|JJ>*<NN.*>.

\subsection{Candidate-Aware Weighting}
We propose a candidate-aware weighting scheme that modulates the decoder's cross-attention based on each candidate's position in the document. For each occurrence of a candidate $c$, we compute its generation probability $p_c$ from the decoder output logits, while applying a candidate-specific weighting vector to the cross-attention at every layer and head. This approach does not introduce any additional inference overhead compared to standard prompt-based methods. The weighting function can be defined separately for each layer, but we omit the layer index $l$ in our notation for clarity.
\subsubsection{\textbf{Vanilla Gaussian Weighting}}
To overcome the limitations of PromptRank in modeling local context, we introduce a candidate-centered Gaussian weighting mechanism for the decoder cross-attention. For each candidate $c$, we construct a vanilla Gaussian Weighting vector $w_{\text{vanilla},c} \in \mathbb{R}^{N}$ over the $N$ document tokens, assigning higher weights to tokens near the candidate’s position and lower weights to more distant tokens:
\begin{equation}
w_{\text{vanilla}, c} = \exp \left( - \frac{(x - \mu_c)^2}{2 \sigma_{l}^2} \right),\quad x = 1, \ldots, N,
\end{equation}
where $x$ denotes the index of a document token and $\mu_c$ denotes the position of candidate $c$ in the document. The standard deviation $\sigma_l$ is defined as $\sigma_l = \sigma_0 + \kappa \cdot l$, with $\sigma_0$ and $\kappa$ as hyperparameters that control the initial spread and its layer-wise growth, respectively. 
This enables the weighting to gradually transition from emphasizing local context in lower layers to capturing more global context in higher layers.

While this improves sensitivity to local context, it may still miss important global or related information. To address this, \algname{} incorporates two additional weighting schemes to aggregate each candidate’s immediate surroundings with related contexts, enabling importance to be assessed across interconnected contexts.
\subsubsection{\textbf{Semantic-aware Gaussian Mixture Weighting}}
To further enhance context modeling, we introduce a semantic-aware Gaussian mixture weighting mechanism. For each candidate $c$, we first compute its embedding $e_c$ by mean-pooling the contextual token embeddings from the encoder of a PLM over its span. We then select the set $C_{\mathrm{sem}}$ of the $k_{\mathrm{sem}}$ most semantically similar candidates to $c$ using cosine similarity:
\begin{equation}
C_{\text{sem}} = \operatorname{Top}\text{-}k_{\text{sem}}\left( \left\{ \cos(e_c, e_{c_j}) \mid c_j \in C,\, c_j \neq c \right\} \right),
\end{equation}
where $C$ is the set of all candidates. And then, to integrate the local contexts of both the candidate $c$ and its semantically related candidates, we define a Gaussian mixture-based weighting:
\begin{equation}
w_{\text{mixture}, c} = \pi_{c} \cdot \mathcal{N}(x; \mu_c, \sigma_l^2) \;+\; \sum_{c_j \in C_{\text{sem}}} \pi_{c_j} \cdot \mathcal{N}(x; \mu_{c_j}, \sigma_j^2),
\end{equation}
where $\pi_c$ and $\pi_{c_j}$ are non-negative mixture coefficients that sum to 1, and $\mu_{c_j}$, $\sigma_j^2$ denote the position and variance of $c_j$ in the document. This mixture Gaussian weighting enables the model to attend not only to the local context of the candidate itself but also to the contexts of its most semantically similar candidates.

\subsubsection{\textbf{Attention-aligned Gaussian Weighting}}
The row-wise Vanilla Gaussian weighting in cross-attention may suppress document tokens that receive strong attention from the candidate span, thereby reducing the model’s ability to capture salient contextual information. To address this, we introduce an attention-aligned weighting mechanism that incorporates alignment information from the cross-attention patterns. Specifically, for each attention head $h$, we use the pre-softmax cross-attention matrix $S^{(h)} \in \mathbb{R}^{M \times N}$, where $M$ and $N$ are the numbers of template and document tokens, respectively. For each document token $j$, we compute an aggregate alignment score $[v^{(h)}]_j  \in \mathbb{R}^{N}$ by summing the attention weights over the candidate’s span in the template:
\begin{equation}
[v^{(h)}]_j  = \sum_{i = k}^{k + \lvert c \rvert - 1} S_{ij}^{(h)},
\end{equation}
where $k$ is the starting row index of the candidate span in the template and $\lvert c \rvert$ is the length of the candidate. This score is computed separately for each head, since different attention heads tend to specialize in attending to different sets of document tokens, capturing complementary aspects of the candidate’s contextual relevance.

We then select the top-$k_{\mathrm{attn}}$ document tokens with the highest $[v^{(h)}]_j$, denoted as $I_{k_{\mathrm{attn}}}^{(h)}$, which represent the most strongly aligned document tokens for the candidate.

\subsubsection{\textbf{Final Candidate-aware Weighting}}
To construct the final weight vector, we combine the Semantic-aware Gaussian mixture weight $w_{\text{mixture},c}$ with the set of document tokens identified by $I_{k_{\text{attn}}}^{(h)}$. Specifically, for each attention head $h$, the final weight vector $w^{(h)}_{\text{final},c} \in \mathbb{R}^{N}$ is defined as:
\begin{equation}
[{w}_{\text{final}, c}^{(h)}]_j = \begin{cases} 1, & \text{if } j \in I_{k_{\text{attn}}}^{(h)} \\ {[ {w}_{\text{mixture}, c} ]_j}, & \text{otherwise} \end{cases}
\end{equation}
We then expand $w^{(h)}_{\text{final}, c}$ to match the number of template tokens, resulting in a weight matrix $W^{(h)}_{\mathrm{final},c} \in \mathbb{R}^{M \times N}$. This matrix is applied element-wise to the cross-attention matrix at each layer:
\begin{equation}
A^{(h)} = \operatorname{softmax}\left( \frac{S^{(h)} \odot W^{(h)}_{\mathrm{final},c}}{\sqrt{d_N}} \right),
\end{equation}
where $d_N$ is the dimension of the document embeddings.

However, since $S^{(h)}$ contains both positive and negative values, directly multiplying negative logits by the weight in [0, 1] can inadvertently amplify the influence of low-weight tokens, contrary to the intended down-weighting effect. Alternatively, dividing negative logits by the weight excessively stretches their range, resulting in highly peaked softmax distribution that may overemphasize specific local contexts. To address this, we stabilize the attention by multiplying negative logits by $(1-\text{weight})$:
\begin{equation}
S^{(h)}_{\mathrm{adj},\,i,\,j} =
\begin{cases}
    S^{(h)}_{i,\,j} \cdot W^{(h)}_{\mathrm{final},\,c,\,i,\,j} & \text{if } S^{(h)}_{i,\,j} \geq 0 \\[10pt]
    S^{(h)}_{i,\,j} \cdot \left(1 - W^{(h)}_{\mathrm{final},\,c,\,i,\,j}\right) & \text{if } S^{(h)}_{i,\,j} < 0
\end{cases}
\end{equation}
For all $S^{(h)}_{i,j}$, we have 
$\partial S^{(h)}_{\mathrm{adj}, i, j} / \partial W \geq 0$, ensuring monotonicity with respect to the importance weight regardless of the sign of the logit. Moreover, the adjusted logits remain bounded within the original range ($S^{(h)}_{\text{adj}} \in [\min(0, S^{(h)}), \max(0, S^{(h)})]$), preventing numerical instability while preserving the intended reweighting effect. The resulting attention map is applied to the value vectors to produce attention-weighted representations, which are then propagated to subsequent layers, allowing both global and local contextual signals to influence the candidate’s generation probability $p_c$ in Eq. \ref{eq:gen_prob}.

\subsection{Multi-Granular Attention}

\subsubsection{\textbf{Self-Attention Map Extraction}}
We obtain the self-attention maps (SAMs) from each layer of the encoder of a PLM. While SAMs naturally capture token-level correlations that reflect the global context of the document, we further aim to incorporate the importance of each candidate within local contexts. To this end, we segment each document into overlapping window blocks, allowing us to focus on local semantic regions as well as the entire document. Each window block and full document are separately fed into the encoder to extract their respective SAMs.
Notably, our approach independently evaluates each candidate occurrence in both global and local attention spaces based on its specific position in the document, rather than relying solely on aggregated or global information. This allows identical terms appearing in different contexts to receive distinct scores.

 Additionally, we empirically select the optimal layer via layer-wise analysis, then average the attention weights across all heads within the selected layer to obtain a robust SAM for scoring. 

\subsubsection{\textbf{Global Attention Score Calculation}}
The global attention score $G_c$ for each occurrence of candidate $c$ is calculated by aggregating attention weights from all tokens to the candidate span $\lvert c \rvert$ from all tokens in the document. The self-attention map $A\in \mathbb{R}^{N \times N}$ is a matrix that represents attention weights between all tokens in a document of length $N$. Therefore, $G_c$ can be computed as a column-wise sum over the SAM.
\begin{equation}
\mathrm{G_c} = \frac{1}{\lvert c \rvert} \sum_{j \in I_{c}} \sum_{i=1}^N \mathrm{A}_{i,j},
\end{equation}
where $I_c$ is the set of token indices for candidate $c$ and $A_{i,j}$ denotes the attention received by token $j$ from token $i$.

\subsubsection{\textbf{Local Attention Score Calculation}}
To assess local contextual importance, we segment the document into overlapping windows (blocks) of consecutive sentences, each centered at sentence $s_k$. For each window block $b_k = \{{s_{k-w}, \ldots, s_k, \ldots, s_{k+w}}\}$ (where $w$ is the window size), we extract the corresponding local SAM $A^{\text{local}}_{b_k}\in \mathbb{R}^{w' \times w'}$, where $w'$ denotes the number of tokens in block $b_k$. We then compute a local attention score $L_c$ for each occurrence of candidate $c$ that falls within the scoring range of $b_k$. The scoring range varies across blocks: in the first block, candidates located at or before the center sentence are considered; in the intermediate blocks, only candidates within the center sentence are evaluated; and in the final block, candidates appearing at or after the center sentence are scored.

The local score $L_c$ for candidate $c$ is computed as:
\begin{equation}
L_c = \mathbb{I}\{c \in b_k\} \cdot \frac{1}{\lvert c \rvert} \sum_{j \in I_c} \sum_{i=1}^{w'} A_{i, j, b_k}^\text{local},
\end{equation}
where $\mathbb{I}\{c \in b_k\}$ is an indicator function that equals $1$ if candidate $c$ is within the scoring range of block $b_k$, and $0$ otherwise. This local scoring mechanism provides a fine-grained assessment of candidate importance within its local context, complementing the global score and enabling a more context-sensitive evaluation.

The final attention-based score for each candidate occurrence is computed as the product of its global and local scores:
\begin{equation}
AS_c = G_c \times L_c.
\end{equation}
This integration of global and local contextual signals enables more precise and context-aware keyphrase extraction.

\subsection{Final Candidate Scoring}
In the final step, we aggregate the generation-based and attention-based scores to compute the final candidate score $FS_c$. Both $p_c$ and $AS_c$ capture the global and local importance of each candidate, but from different modeling perspectives: generation-based and attention-based, respectively. By integrating these complementary scores, we obtain a robust measure of candidate salience from both semantic and contextual signals. The final score is computed as:
\begin{equation}
FS_c= \frac{p_c}{1+\log(1+\lambda\cdot AS_c)},
\end{equation}
where the balancing parameter $\lambda$ allows us to control the contribution of attention-based importance. The log normalization in the denominator mitigates the effect of large attention values, resulting in stable and effective candidate ranking.

\subsection{Extension to Decoder-only Models}
The aforementioned framework on an encoder-decoder architecture can be naturally extended to decoder-only architectures, which lack cross-attention and process all input via masked self-attention.

\subsubsection{\textbf{Candidate-Aware Weighting}}
To apply candidate-aware weighting, we concatenate the document and the candidate-containing template as a single input sequence. The pre-softmax self-attention matrix $S^{(h)}$ then covers both document and template tokens, with shape $(N+M) \times (N+M)$, where $N$ and $M$ are the lengths of the document and template, respectively.

\vspace{0.1cm}
\noindent\emph{\underline{Semantic-aware Gaussian Mixture Weighting}}: To identify semantically related candidates ($C_{\mathrm{sem}}$) in decoder-only models, candidate and document embeddings can be derived by using an off-the-shelf pretrained encoder. 

\vspace{0.1cm}
\noindent\emph{\underline{Attention-aligned Gaussian Weighting}}: The aggregate alignment score for the $j$-th document token in head $h$ is computed by summing the relevant rows of the self-attention matrix corresponding to the candidate's span in the template:
\begin{equation}
[v^{(h)}]_j  = \sum_{i = N+k}^{N+k + \lvert c \rvert - 1} S_{ij}^{(h)}, \quad j = 1, \dots, N,
\end{equation}
where $N + k$ is the starting row index of the candidate in the concatenated input, and $j$ is limited to the range of document tokens. 

\vspace{0.1cm}
\noindent\emph{\underline{Final Weight Matrix}}: The candidate-aware weight matrix $W^{(h)}_{\mathrm{final},c} \in \mathbb{R}^{N+M \times N+M}$ is structured as:
\begin{equation}
W^{(h)}_{\mathrm{final},c} =
\begin{bmatrix}
\mathbf{1}_{N \times N} & \mathbf{1}_{N \times M} \\
w^{(h)}_{\mathrm{final},c} & \mathbf{1}_{M \times M}
\end{bmatrix},
\end{equation}
where $\mathbf{1}$ denotes an all-ones matrix and $w^{(h)}_{\mathrm{final},c}$ is the candidate-aware weight vector for document tokens.

\subsubsection{\textbf{Multi-Granular Attention}}
In decoder-only models, masked self-attention restricts each token to attend only to previous tokens, resulting in a strictly lower triangular attention map, with nonzero values only on and below the diagonal. The attention-based importance score $AS_c$ is computed in the same way as for encoder-decoder models, by aggregating attention weights (e.g., via column-wise sum) over the candidate span.

\subsection{Complexity Analysis}
Given a fixed-cost backbone PLM, \algname{} has linear complexity $O(D \cdot n + D \cdot B)$ in the number $D$ of documents in a corpus, decomposed into (i) candidate-level inference with $O(D \cdot n)$ and (ii) self-attention map extraction with $O(D \cdot (B + 1))$, where $n$ and $B$ denote the number of candidates and overlapping blocks per document, respectively.
\section{Experiments}

\begin{table}[t]
\small
\centering
\caption{Statistics of datasets.}
\vspace{-0.3cm}
\label{tab:dataset_stats}
\resizebox{\columnwidth}{!}{%
\begin{tabular}{lcccc}
\toprule
\textbf{Dataset} & \textbf{DUC2001} & \textbf{WikiHow} & \textbf{NUS} & \textbf{SemEval2017} \\
\midrule
\# Documents & 308 & 500 & 211  & 493 \\
Avg. Doc Length & 725 & 996 & 7702 & 170 \\
Avg. \# Sentences & 34 & 51& 425 & 7 \\
Avg. \# Keyphrases & 8 & 16 & 12 & 17\\
Absent  Keyphrases (\%) & 2.2 & 35.8 & 15.9 & 0.3 \\
Noun Phrases (\%) & 91.6 & 96.4 & 82.4 & 62.8 \\
Topic Drift Score & 0.47 & 0.50 & 0.41  & 0.38 \\
\bottomrule
\end{tabular}
}
\end{table}


\begin{table*}[t]
\centering
\caption{Performance comparison (F1@\textit{k}) of keyphrase extraction methods on four datasets. Statistically significant improvements over all baselines are marked with * (paired t-test, $p < 0.05$).}
\vspace{-0.2cm}
\label{tab:keyphrase-comparison}
\begin{adjustbox}{max width=\textwidth}
\begin{tabular}{
p{3.2cm} p{3.8cm} ccc ccc ccc ccc }
\toprule
\textbf{Method Type} & \textbf{Method} & \multicolumn{3}{c}{\textbf{DUC2001}} & \multicolumn{3}{c}{\textbf{WikiHow}} & \multicolumn{3}{c}{\textbf{NUS}} & \multicolumn{3}{c}{\textbf{SemEval2017}} \\
\cmidrule(lr){3-5} \cmidrule(lr){6-8} \cmidrule(lr){9-11} \cmidrule(lr){12-14}
& & F1@5 & F1@10 & F1@15 & F1@5 & F1@10 & F1@15 & F1@5 & F1@10 & F1@15 & F1@5 & F1@10 & F1@15 \\
\midrule

Statistic-based & YAKE  & 11.99 & 14.18 & 14.28 &\textbf{18.61} & \underline{20.89} & 21.12 & 7.85 & 11.05 & 13.09 & 11.84 & 18.14 & 20.55\\

\hdashline
\multirow{4}{*}{\centering Graph-based} & TextRank & 11.02 & 17.45 & 18.84 & 2.91 & 5.27 & 6.70 & 1.80 & 3.02 & 3.53 & 16.43 & 25.83 & 30.50 \\
 & SingleRank & 19.14 & 23.86 & 23.43 & 7.59 & 11.77 & 13.51 & 2.98 & 4.51 & 4.92  & 18.23 & 27.73 & 31.73\\
 & TopicRank & 19.97 & 21.73 & 20.97 & 16.77 & 18.47 & 18.05 & 4.54 & 7.93 & 9.37 &  17.10 & 22.62 & 24.87\\
 & MultipartiteRank & 21.70 & 24.10 & 23.62 & \underline{17.89} & 20.44 & 20.40 & 6.17 & 8.57 & 10.82  
 & 17.39 & 23.73 & 26.87\\
 
\hdashline
\multirow{3}{*}{\centering Embedding-based} & EmbedRank (BERT)  & 8.12 & 11.62 & 13.58 & 2.35 & 4.29 & 5.84 & 3.75 & 6.34 & 8.11 & 20.03 & 31.01 & 36.72\\
& JointGL (BERT) &26.03	&28.76	&29.07 &13.02	&16.24	&17.84	&10.82	&13.86	&16.06	 & 18.05 &26.00	&29.27\\
 & MDERank (BERT) & 13.05 & 17.31 & 19.13 & 15.25 & 18.80 & 19.06 & 15.24 & 18.33 & 17.95   & 22.81 & 32.51 & 37.18\\

\hdashline
\multirow{2}{*}{\centering Attention-based} & SAMRank (Llama3)  & 28.26 & 30.47 & 30.20 & 8.16 & 10.18 & 11.49& 19.85 & 21.31 & 21.27  & 24.74 & 33.51 & 37.01\\
 & Attention-Seeker (Llama3)  & 29.71 & 31.75 & 31.11 & 8.65 & 11.22 & 12.42& \underline{20.55} & 22.01 & 22.02 & 25.40 & 34.53 & 38.50 \\

\hdashline
\multirow{5}{*}{\centering Prompt-based}  & Empirical (Llama3)  & \underline{30.43} & 32.31 & 29.76 & 10.97 & 15.08 & 16.25&  20.53 & 23.19 & 22.61  & 23.84 & 33.06 & 37.53\\ 
& PromptRank (T5)  & 23.71 & 28.38 & 28.43 & 12.48 & 17.21 & 19.04 & 15.81 & 18.99 & 19.59  & 27.07 & 37.83 & \underline{41.82}\\
 & \cellcolor{lightgray!15}\textbf{\algname{} (T5)} & \cellcolor{lightgray!15}26.99 & \cellcolor{lightgray!15}31.95 & \cellcolor{lightgray!15}31.97 & \cellcolor{lightgray!15}15.54 & \cellcolor{lightgray!15}20.37 & \cellcolor{lightgray!15}\underline{22.26}& \cellcolor{lightgray!15}16.84 & \cellcolor{lightgray!15}19.95 & \cellcolor{lightgray!15}20.41  & \cellcolor{lightgray!15}27.02 & \cellcolor{lightgray!15}37.83 & \cellcolor{lightgray!15}\underline{41.82}\\
 & PromptRank (Gemma2) & 29.28 & \underline{33.54} & \underline{33.59} & 14.90 & 19.96 & 21.43 &
 \textbf{21.62} & \underline{23.85} & \underline{23.32}   & \underline{27.31} & \underline{38.02} & 41.60\\
 & \cellcolor{lightgray!15}\textbf{\algname{} (Gemma2)} & \cellcolor{lightgray!15}\textbf{32.77*} & 
 \cellcolor{lightgray!15}\textbf{36.86*} & 
 \cellcolor{lightgray!15}\textbf{36.00*} & 
 \cellcolor{lightgray!15}16.59 & 
 \cellcolor{lightgray!15}\textbf{21.84*} & 
 \cellcolor{lightgray!15}\textbf{23.47*} &
 \cellcolor{lightgray!15}\textbf{21.62} & 
 \cellcolor{lightgray!15}\textbf{24.15} & 
 \cellcolor{lightgray!15}\textbf{24.28*} & 
 \cellcolor{lightgray!15}\textbf{27.39} & 
 \cellcolor{lightgray!15}\textbf{38.36} & 
 \cellcolor{lightgray!15}\textbf{41.84} \\
\bottomrule
\end{tabular}
\end{adjustbox}
\end{table*}

\subsection{Datasets and Evaluation Metrics}

We evaluated \algname{} on four real-world datasets. DUC2001~\cite{duc} comprises news articles, NUS~\cite{nus} provides full scientific papers, and SemEval2017~\cite{semeval2017} consists of paper abstracts. In addition to these widely used benchmarks, WikiHow~\cite{wikihow} dataset covers instructional articles on a wide spectrum of real-world topics, such as cooking, health, and technology. While existing benchmarks are domain-specific and often exhibit positional bias, WikiHow provides section-level gold summaries where extracted keyphrases are widely distributed, enabling broader and more robust evaluation.

Table~\ref{tab:dataset_stats} summarizes dataset statistics and characteristics. We measure topical diversity by segmenting each document into three-sentence units and computing the average pairwise cosine similarity of their embeddings. For each dataset, we report the mean topic drift score across all documents, defined as one minus the average similarity. DUC2001 and WikiHow exhibit higher topic drift than the scientific datasets NUS and SemEval2017, indicating greater topical variation in news and instructional texts.

Following previous works, we evaluate keyphrase extraction performance using F1@$k$ where $k$ is the top 5, 10, and 15 ranked candidates. All baselines consider noun phrases evaluated against gold keyphrases, including those absent from the original documents. For candidates appearing multiple times in different contexts, we retain only the occurrence with the highest score. 

\begin{table*}[t]
\small
\centering
\caption{Ablation study results.}
\vspace{-0.2cm}
\label{tab:ablation_results}
\begin{adjustbox}{max width=\textwidth}
\begin{tabular}{l ccc ccc ccc ccc}
\hline
\textbf{Method} & \multicolumn{3}{c}{\textbf{DUC2001}} & \multicolumn{3}{c}{\textbf{WikiHow}} & \multicolumn{3}{c}{\textbf{NUS}} & \multicolumn{3}{c}{\textbf{SemEval2017}} \\
\cmidrule(lr){2-4} \cmidrule(lr){5-7} \cmidrule(lr){8-10} \cmidrule(lr){11-13}
 & F1@5 & F1@10 & F1@15 & F1@5 & F1@10 & F1@15 & F1@5 & F1@10 & F1@15 & F1@5 & F1@10 & F1@15 \\
\hline
\textbf{\algname{}} (default)                   & 
\textbf{26.99} & 
\textbf{31.95} & 
\textbf{31.97} & 
\textbf{15.54} & 
\textbf{20.37} & 
\textbf{22.26} &
\textbf{16.84} & \textbf{19.95} & \underline{20.41} & 27.02 & 37.83 & 41.82 \\
\hline
~\textbf{Candidate-Aware Weighting} only  & 25.85 & 29.35 & 29.05 & 13.67 & 17.85 & 19.78 & 
\underline{16.78} & \underline{19.91} & \textbf{20.48}  & 27.04 & \underline{37.94} & \underline{41.95}\\
~~ $w_{\text{vanilla}, c}$ $+$ Attention-aligned     & 25.55 & 29.43 & 28.94 & 13.40 & 17.58 & 19.48& 16.38 & 19.12 & 19.77  & \textbf{27.26} & \textbf{37.99} & \textbf{42.08}\\
~~ $w_{\text{vanilla}, c}$ $+$ Semantic-aware Mixture    & 23.41 & 27.34 & 28.40 & 13.05 & 17.55 & 19.64 & 15.36 & 18.51 & 19.69  & 26.60 & 37.57 & 41.49\\
~~ $w_{\text{vanilla}, c}$ (vannila weighting only)     & 23.31 & 27.73 & 28.15 & 13.03 & 17.46 & 19.39& 15.59 & 18.46 & 19.09    & 26.70 & 37.50 & 41.59\\

\hline
~\textbf{Multi-Granular Attention} only        & \underline{26.64} & 31.34 & \underline{31.58} & \underline{15.36} & \underline{19.99} & \underline{21.88}& 16.10 & 19.16 & 19.59 & 27.10 & 37.77 & 41.74\\
~~ $G_c$ (global context only)            & 26.25 & \underline{31.52} & 31.07 & 14.49 & 19.55 & 21.25 & 15.98 & 19.08 & 19.55 & 27.11 & 37.80 & 41.69 \\
~~ $L_c$ (local context only)             & 25.05 & 29.72 & 30.02 & 14.27 & 19.12 & 20.75& 15.98 & 18.77 & 19.69 & \underline{27.19} & 37.75 & 41.70 \\
\hline
PromptRank             & 23.71 & 28.38 & 28.43 & 12.48 & 17.21 & 19.04& 15.81 & 18.99 & 19.59  & 27.07 & 37.83 & 41.82\\
\hline
\end{tabular}
\end{adjustbox}
\end{table*}

\subsection{Baselines and Implementation Details}

We compare \algname{} against a comprehensive set of baselines, including: statistics-based (YAKE ~\cite{yake}), graph-based (TextRank~\cite{textrank}, SingleRank ~\cite{duc}, TopicRank~\cite{topicrank}, MultipartiteRank ~\cite{multirank}), embedding-based (EmbedRank ~\cite{embedrank}, JointGL ~\cite{jointgl}, MDERank ~\cite{mderank}), attention-based (SAMRank ~\cite{samrank}, Attention-Seeker ~\cite{attention-seeker}), and prompt-based (PromptRank ~\cite{kong2023promptrank}, Empirical ~\cite{empirical}) methods. Consistent with the other baselines, Empirical identifies noun phrases as candidates and utilizes role-based prompting to select the top-$k$ candidates from the provided set. We follow the same PLM backbones (BERT, Llama3, and T5-base), length penalty ($\alpha$), and the maximum token length of 512, as used in the prior works. Additionally, to enable direct comparison in the decoder-only setting, we adapt both PromptRank and \algname{} to use Gemma2-9B~\cite{gemma2}. For all baselines, NLTK’s PorterStemmer is used for stemming.

For all datasets, \algname{} sets a fixed context size with initial distribution spread $\sigma_0=0.3$, mixture weights $\pi_c=0.7$ and $\pi_{c_j}=0.1$, and a window $w=1$. The numbers $k_{sem}$ of semantically similar candidates and $k_{attn}$ of highly aligned document tokens are set to 3 and 15, respectively. The two key hyperparameters, a distribution scaling factor $\kappa$ and an attention score balancing factor $\lambda$, are tuned based on validation performance, of which sensitivity is analyzed. The detailed settings are provided in the source code repository.


\subsection{Overall Results}
Table~\ref{tab:keyphrase-comparison} reports F1@5, F1@10, and F1@15 scores for \algname{} and baselines across four datasets using both the encoder-decoder (T5) and the decoder-only (Gemma2) backbones. The overall results show the effectiveness of \algname{} in modeling local context to enhance keyphrase extraction across different model architectures and document types.

Specifically, \algname{} (Gemma2) achieves the highest performance for most cases, by a large margin on DUC2001 and WikiHow, which feature longer and more topically diverse corpora. On average, \algname{} outperforms the state-of-the-art PromptRank (Gemma2) by margins of 6.52\%, 5.59\%, and 5.35\% at F1@5, F1@10, and F1@15, respectively.
On NUS, where documents are long but less topically diverse, improvements are moderate, while for SemEval2017, which consists of short, topically focused texts, both \algname{} and PromptRank perform similarly, suggesting that prompt-based scoring suffices in such settings. With T5, \algname{} similarly achieves strong results, outperforming all baselines except PromptRank (Gemma2) on three of four datasets, excluding NUS. The most substantial gains again appear on datasets with higher topic drift, whereas performance on more topically coherent collections is comparable to existing methods. Furthermore, while Empirical shows that diverse prompting strategies improve zero-shot keyphrase extraction, our results suggest that prompt design alone is insufficient to capture the multi-granular topic saliency of the candidate.

\subsection{Ablation Study}
We conduct an ablation study on two key components, multi-granular attention and candidate-aware weighting, of \algname{} with a default encoder-decoder model. Table~\ref{tab:ablation_results} shows the evaluation results, along with PromptRank as the reference baseline. 

For the candidate-aware weighting component, our results demonstrate that applying the final weighting improves performance over all other variants across all datasets. These findings highlight that weighting candidates in a way that reflects both their immediate context and their strong ties to other meaningful content in the document leads to more accurate keyphrase selection. However, on SemEval2017, integrating Semantic-aware Mixture with vanilla weighting leads to a performance drop, as the additional context in short documents dilutes the focus on local cues.

For the multi-granular attention component, our results show that it leads to notable performance improvements on datasets with high topic drift, such as DUC2001 and WikiHow, especially when both local and global signals are combined. This underscores the value of multi-level contextual modeling. In contrast, its effect is limited or even negative on more homogeneous datasets such as NUS and SemEval2017, where generation-based candidate scoring alone is sufficient and adding multi-granular attention may introduce noise. Overall, the full model achieves the best results on DUC2001 and WikiHow, while candidate-aware weighting alone performs best on NUS and SemEval2017.

\begin{figure}[t]
    \centering
    \includegraphics[width=0.8\linewidth]{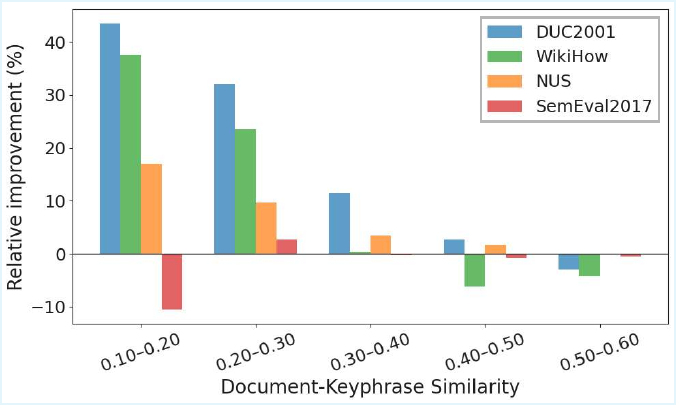}
    \vspace{-0.2cm}
    \caption{Relative improvement of \algname{} over PromptRank on Top-15 recall of gold keyphrases across document–keyphrase similarity bins.}
    \vspace{-0.2cm}
    \label{fig:recall}
\end{figure}
\subsection{Properties of Extracted Keyphrases}

To assess the ability of \algname{} to extract gold keyphrases from semantically diverse document regions, we measure the recall of gold keyphrases in the top-15 extracted candidates across document-keyphrase similarity bins. For each bin, recall is defined as the proportion of gold keyphrases correctly retrieved among the top-15 candidates. We compare PromptRank and \algname{}, and report the relative improvement of \algname{} over PromptRank on Top-15 recall across similarity bins.

As shown in Figure~\ref{fig:recall}, PromptRank achieves higher recall for keyphrases with high similarity to the document, indicating a bias toward globally salient terms. In contrast, \algname{} achieves higher recall in lower similarity bins and remains robust across higher similarity bins, effectively capturing keyphrases from both global and local contexts. This effect is especially pronounced on DUC2001 and WikiHow, where our approach substantially outperforms PromptRank in low-similarity regions, highlighting its advantage for documents with high topic drift. On NUS, \algname{} still improves recall in low-similarity bins, though the gap is smaller due to lower topic drift in scientific articles. On SemEval2017, however, our method shows slightly lower recall than PromptRank in the low-similarity bins, likely because the documents are relatively short and homogeneous, limiting the advantage of our approach.

\subsection{Qualitative Case Study}
\begin{figure}[t]
  \centering
  \begin{subfigure}[b]{0.45\textwidth}
    \centering
    \includegraphics[width=\textwidth]{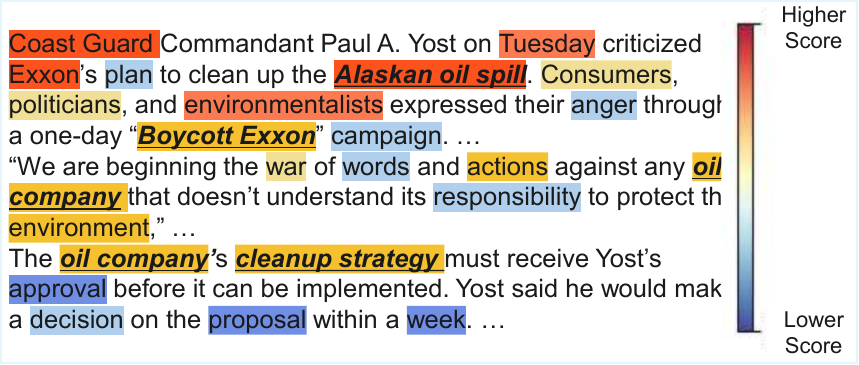}
    \vspace{-0.6cm}
    \caption{PromptRank.}
    \label{fig:comp_a}
  \end{subfigure}
  \hfill
  \begin{subfigure}[b]{0.45\textwidth}
    \centering
    \includegraphics[width=\textwidth]{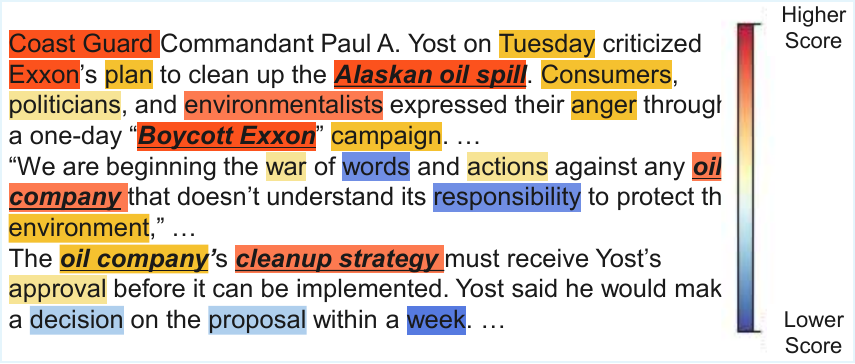}
    \vspace{-0.6cm}
    \caption{\algname{}.}
    \label{fig:comp_b}
  \end{subfigure}
  \vspace{-0.2cm}
  \caption{Case study on an example document with the same keyphrases scored by PromptRank and \algname{}, respectively. Each keyphrase score is highlighted using heat maps.}
  \label{fig:comp}
\end{figure}
To demonstrate the effectiveness of \algname{}, we present a case study using a document from the DUC2001 dataset. Figure~\ref{fig:comp} compares candidate scores assigned by PromptRank and \algname{}, visualized as normalized heat maps, where warmer colors indicate higher scores and gold keyphrases are marked in bold italics. Overall, \algname{} more accurately assigns higher scores to gold keyphrases compared to PromptRank. Notably, while PromptRank assigns identical scores to repeated terms such as \texttt{oil company}, \algname{} differentiates their importance by considering the local context. Specifically, \texttt{oil company} is more relevant to the second paragraph, which centers on public anger toward \texttt{oil company}, whereas the third discusses \texttt{cleanup strategy}. Considering these contexts, \algname{} assigns a higher score to \texttt{oil company} when it appears in the second paragraph, effectively identifying it as a salient keyphrase in the appropriate local context.

\subsection{Sensitivity Analysis}

\begin{figure}[t]
    \centering
    \includegraphics[width=0.8\linewidth]{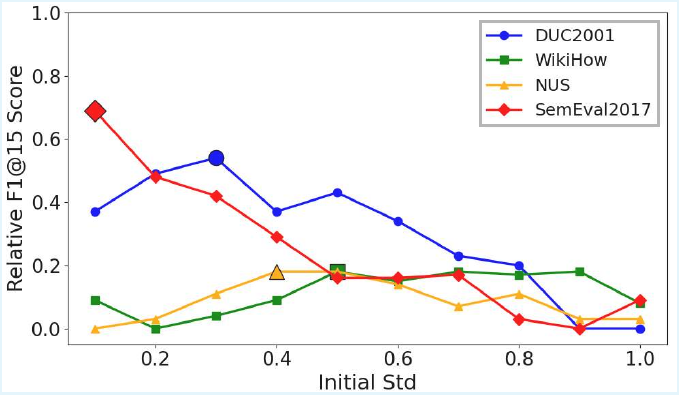}
    \vspace{-0.2cm}
    \caption{Performance with different $\sigma_0$ values. Y-axis indicates the relative F1@15 score, defined as the F1@15 at each setting minus the minimum F1@15 across all settings.}
    \label{fig:relative_f1_weight}
\end{figure}

\subsubsection{\textbf{Initial Distribution Spread $\sigma_0$}}
We analyze the effect of the initial standard deviation $\sigma_0$ by varying it from 0.1 to 1.0 (step size: 0.1) and, for each dataset, measuring the relative change in F1@15 after normalizing the scores such that the minimum value is set to zero. $\sigma_0$ controls the spread of the weighting, determining how much local context is considered during decoding. As shown in Figure~\ref{fig:relative_f1_weight}, performance tends to improve as $\sigma_0$ increases, up to a certain point, after which it declines, highlighting the need to balance local and global context. For short-document datasets (SemEval2017), optimal performance is achieved at $\sigma_0=0.1$, while longer-document datasets (DUC2001,  WikiHow, NUS) are robust over a wider range and peak between 0.3 and 0.5. While trends differ across datasets, the overall performance differences are not substantial, confirming the robustness of \algname{}. 

\begin{table}[t]
\centering
\caption{Effect of window size on F1@15 score.}
\vspace{-0.2cm}
\label{tab:window_size}
\begin{tabular}{c|ccc}
\hline
Window Size & DUC2001 & WikiHow & NUS \\
\hline
1 & 31.97 & 22.26 & 20.41\\
2 & 32.09 & 22.56 & 20.41\\
3 & 32.28 & 22.51 & 20.37 \\
\hline
\end{tabular}
\end{table}

\subsubsection{\textbf{Window Size $w$}}
To investigate the effect of window size on performance, we evaluated the F1@15 scores while varying the window size parameter ($w \in \{1, 2, 3\}$). We excluded SemEval2017 from this analysis due to its short document lengths. As shown in Table~\ref{tab:window_size}, each dataset exhibits different optimal values, but performance is not highly sensitive. These results suggest that assessing candidate importance within a local context that approximates paragraph- or subtopic-level boundaries is particularly effective even with a small window size. For simplicity and generality across diverse datasets, we fix the window size to $w=1$.

\subsubsection{\textbf{Attention Score Balancing Factor $\lambda$}}

\begin{figure}[t]
    \centering
    \includegraphics[width=0.8\linewidth]{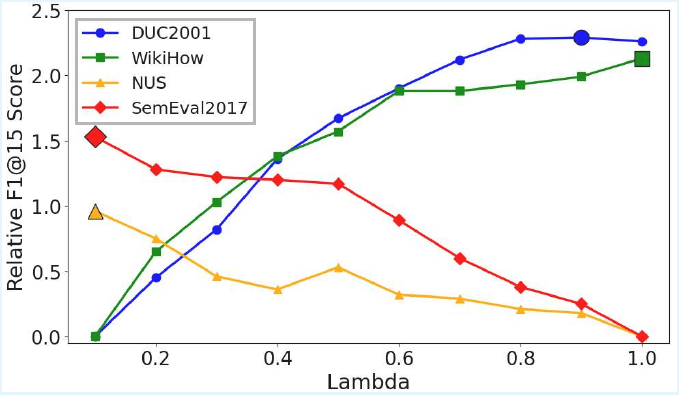}
    \vspace{-0.2cm}
    \caption{Performance with different $\lambda$ values. Y-axis indicates the relative F1@15 score, defined as the F1@15 at each setting minus the minimum F1@15 across all settings.}
    \label{fig:relative_f1}
\end{figure}

Figure~\ref{fig:relative_f1} shows the relative change in F1@15 with varied $\lambda$, illustrating how this parameter controls the contribution of attention-based importance and balances generation-based and attention-based signals (Eq. 12). The optimal values of $\lambda$ are $0.9$ for DUC2001, $1$ for WikiHow, and $0.1$ for SemEval2017 and NUS. On topically coherent datasets such as SemEval2017 and NUS, the benefit of multi-granular attention becomes limited. In contrast, for datasets with greater topic drift (DUC2001 and WikiHow), higher values of $\lambda$ improve performance, as attention-based signals provide complementary information beyond generation-based scores.

\begin{figure}[t]
    \centering
    \includegraphics[width=0.8\linewidth]{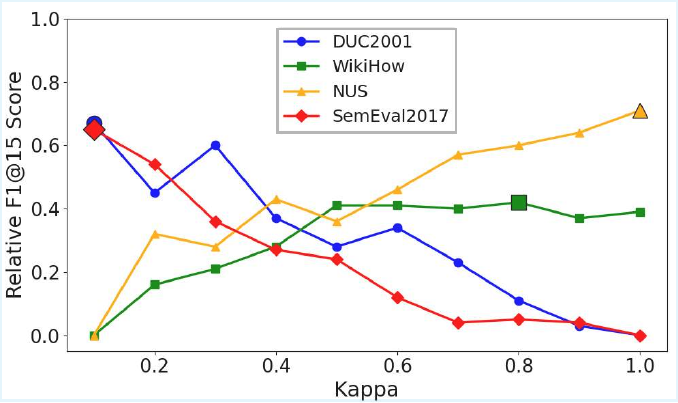}
    \vspace{-0.2cm}
    \caption{Performance with different $\kappa$ values. Y-axis indicates the relative F1@15 score, defined as the F1@15 at each setting minus the minimum F1@15 across all settings.}
    \label{fig:relative_f1_scaling}
\end{figure}

\subsubsection{\textbf{Distribution Scaling Factor $\kappa$}}
The scaling factor $\kappa$ determines how rapidly the weighting expands with layer depth, enabling the model to gradually shift its attention from local to global context during decoding. As shown in Figure \ref{fig:relative_f1_scaling}, the best $\kappa$ settings varied across datasets from $0.1$ for DUC2001 and SemEval2017, $0.8$ for WikiHow, and to $1.0$ for NUS. For DUC2001 and WikiHow, which exhibit greater topic drift, gradual expansion to a broader context proves beneficial, though excessive expansion can dilute relevant local signals. In NUS, where topic drift is low, larger $\kappa$ values are needed to avoid extracting noisy terms from the local context. By contrast, in SemEval2017, the short document length makes minimal scaling sufficient to capture contextual information.

\begin{figure}[t]
    \centering
    \includegraphics[width=0.85\linewidth]{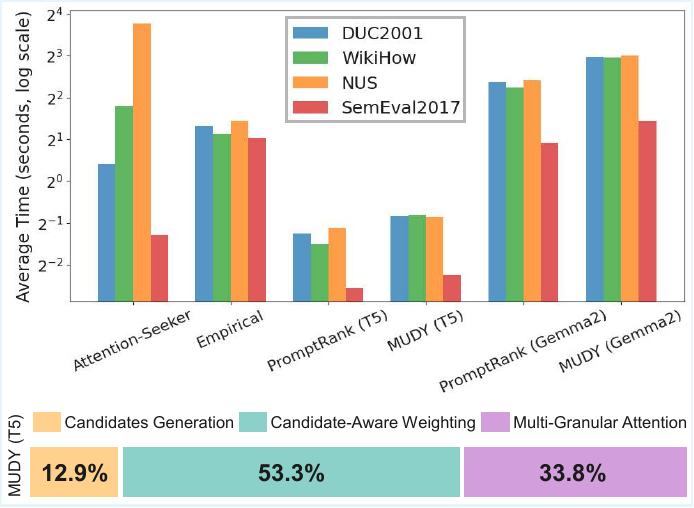}
    \vspace{-0.2cm}
    \caption{Average computation time per document (seconds) of representative PLM-based methods and percentage of total inference time for each step in MUDY.}
    \label{fig:processing_time_total}
\end{figure}


\begin{table*}[t]
\centering
\caption{Performance comparison between GPT-5 and \algname{}.}
\vspace{-0.1cm}
\label{tab:gpt_comparison}
\begin{adjustbox}{max width=\textwidth}
\begin{tabular}{p{3.8cm} ccc ccc ccc ccc}
\toprule
\textbf{Method} 
  & \multicolumn{3}{c}{\textbf{DUC2001}} 
  & \multicolumn{3}{c}{\textbf{WikiHow}} 
  & \multicolumn{3}{c}{\textbf{NUS}} 
  & \multicolumn{3}{c}{\textbf{SemEval2017}} \\
\cmidrule(lr){2-4} \cmidrule(lr){5-7} \cmidrule(lr){8-10} \cmidrule(lr){11-13}
& F1@5 & F1@10 & F1@15 & F1@5 & F1@10 & F1@15 & F1@5 & F1@10 & F1@15 & F1@5 & F1@10 & F1@15 \\
\midrule
GPT-5 (Vanilla)   & 26.57 & 26.27 & 23.75 & 9.98  & 12.23 & 12.82 & \textbf{23.95} & \textbf{25.86} & \underline{23.36} & 22.53 & 31.23 & 36.46 \\
GPT-5 (Candidate) & \textbf{34.12} & \underline{35.51} & \underline{33.96} & 11.91 & 14.43 & 15.52 & \underline{21.62} & 21.54 & 20.15 & 25.19 & 35.85 & 40.37 \\
\midrule
\algname{} (T5)     & 26.99 & 31.95 & 31.97 & \underline{15.54} & \underline{20.37} & \underline{22.26} & 16.84 & 19.95 & 20.41 & \underline{27.02} & \underline{37.83} & \underline{41.82} \\
\algname{} (Gemma2) & \underline{32.77} & \textbf{36.86} & \textbf{36.00} & \textbf{16.59} & \textbf{21.84} & \textbf{23.47} & 21.62 & \underline{24.15} & \textbf{24.28} & \textbf{27.39} & \textbf{38.36} & \textbf{41.84} \\
\bottomrule
\end{tabular}
\end{adjustbox}
\end{table*}

\begin{table*}[t]
\centering
\caption{Performance comparison on large-scale datasets.}
\vspace{-0.1cm}
\label{tab:large_scale}
\begin{tabular}{p{3.8cm} ccc ccc ccc}
\toprule
\textbf{Method} 
  & \multicolumn{3}{c}{\textbf{KPTimes}} 
  & \multicolumn{3}{c}{\textbf{StackExchange}} 
  & \multicolumn{3}{c}{\textbf{KP20k}} \\
\cmidrule(lr){2-4} \cmidrule(lr){5-7} \cmidrule(lr){8-10}
& F1@5 & F1@10 & F1@15 
& F1@5 & F1@10 & F1@15 
& F1@5 & F1@10 & F1@15 \\
\midrule
AttentionSeeker (Llama3)
  &  8.71 &  8.81 &  8.67
  &  7.36 &  8.10 &  8.19
  & \underline{16.05} & 14.73 & 13.22 \\
PromptRank (T5)
  & \underline{14.30} & \underline{14.99} & \underline{13.95}
  & \underline{15.27} & \underline{12.64} & \underline{10.37}
  & 15.75 & \underline{16.02} & \underline{14.64} \\
MUDY (T5)
  & \textbf{18.00} & \textbf{16.91} & \textbf{14.91}
  & \textbf{16.72} & \textbf{13.16} & \textbf{10.56}
  & \textbf{16.32} & \textbf{16.40} & \textbf{14.87} \\
\bottomrule
\end{tabular}%
\end{table*}

\subsection{Processing Time Analysis}

We evaluate the efficiency of \algname{} compared to PLM-based methods, Attention-Seeker (Llama3-8B), Empirical (Llama3-8B), and PromptRank (T5-base and Gemma2-9B) using an NVIDIA B200 GPU. As shown in Figure \ref{fig:processing_time_total}, we observe that processing time is primarily governed by the PLM backbone. While the lightweight encoder-decoder T5 exhibits the highest efficiency, decoder-only models show comparable processing time regardless of the specific method. \algname{} introduces no significant computational overhead compared with existing baselines, demonstrating its practicality.

We further break down the inference time across the three steps of \algname{}, averaged over four datasets. Candidate Generation is the most lightweight stage, followed by Multi-Granular Attention and Candidate-Aware Weighting, both of which constitute the core components responsible for multi-granular contextualization.

\subsection{Extended Evaluation}

\subsubsection{\textbf{Comparison with a Frontier LLM}}
To investigate the competitiveness of \algname{} against large-scale proprietary models, we conduct a comparative analysis with GPT-5~\cite{gpt-5} in a zero-shot keyphrase extraction setting. We employ two prompting strategies: \textbf{Vanilla}, where the model is prompted with role instructions to extract the top 15 keyphrases from a given document, and \textbf{Candidate}, where a candidate list is additionally provided as input and the model is prompted to select the top 15 keyphrases from it.
 
Table~\ref{tab:gpt_comparison} shows that \algname{} (T5) is comparable to GPT-5, and \algname{} (Gemma2) outperforms GPT-5 under both prompting strategies in most cases, especially on DUC2001 and WikiHow. To better understand the impact of candidate lists on GPT-5, we analyze the performance gap between the two prompting strategies. While providing a candidate list generally improves GPT-5's performance, it degrades results on NUS due to its relatively higher absent keyphrase ratio ($\sim$16\%) and lower noun phrase coverage ($\sim$82\%). In this case, constraining the model's generation to a noisier candidate list appears to hinder GPT-5's generative capacity. Notably, \algname{} with a lightweight backbone not only achieves superior performance but also offers a more cost-effective alternative, requiring no API access for resource-constrained applications.
\subsubsection{\textbf{Performance on Large-Scale Benchmarks}}
To further evaluate the generalizability of \algname{}, we conduct experiments on three large-scale datasets spanning diverse domains. KPTimes~\cite{kptimes} and KP20k~\cite{kp20k} each contain approximately 20,000 documents from news articles and paper abstracts, respectively. 
StackExchange ~\cite{stackexchange} consists of 16,000 documents sourced from computer science Q\&A forums, where user-assigned tags serve as keyphrase labels. Since StackExchange provides a tag recommendation system that suggests general, topic-level terms, the keyphrases tend to reflect the global topics of a document rather than specific textual spans.

Table~\ref{tab:large_scale} shows that \algname{} (T5) achieves the best performance across all three datasets, outperforming the latest (AttentionSeeker) and the strongest (PromptRank) baseline. The advantage is particularly pronounced on KPTimes, which features longer documents and high topic diversity. On StackExchange, where keyphrases generally reflect global topics, \algname{} still demonstrates strong performance, owing to its ability to capture global topic saliency. Notably, all three datasets have relatively few ground-truth keyphrases per document (5.15, 2.7, and 5.3 on average for KPTimes, StackExchange, and KP20k, respectively). At F1@15, precision drops sharply as predictions exceed the ground truth, while recall saturates across methods, compressing the performance gap. The advantage of \algname{} is thus more distinctly reflected in F1@5 and F1@10.

\section{Conclusion}
In this paper, we propose an unsupervised keyphrase extraction framework \algname{} that captures multi-granular topic saliency by modeling unique local importance for each candidate occurrence. Prior approaches typically assign a single score to each candidate within a document by focusing on its global relevance to the document, overlooking the influence of distinct local contexts. In contrast, our method leverages candidate-aware weighting in prompt-based generation and integrates multi-granular importance using self-attention maps. This allows \algname{} to extract locally salient keyphrases that are often missed by existing methods focused solely on global relevance. Extensive evaluation on four benchmark datasets with 12 baselines shows that our approach consistently outperforms prior work, especially on long documents with significant topic drift, and achieves higher recall of gold keyphrases from semantically diverse regions of the document. 

\section*{Acknowledgments}
This work was partly supported by the Institute of Information \& Communications Technology Planning \& Evaluation (IITP)-ICT Creative Consilience Program (IITP-2026-RS-2020-II201819), IITP-ITRC (Information Technology Research Center) (IITP-2026-RS-2024-00436857), and the National Research Foundation of Korea (NRF) (RS-2024-00406320 and RS-2026-25494369) funded by the Korea government(MSIT) .

\balance
\bibliography{References}
\end{document}